\documentclass{pasj00}

\usepackage{lscape}

\def\kms{~km~s$^{-1}$}
\def\h2o{H$_2$O}
\def\vlsr{$V_{\mbox{\scriptsize LSR}}$\ }
\def\etal{~et~al.\ }

\def\r22{IRAS\,22555$+$6213}
\def\iras{IRAS\,22198$+$6336}
\def\ga{G108.18$+$05.51}
\def\gb{G108.20$-$00.58}
\def\gc{G108.47$+$02.81}
\def\gd{G108.59$+$00.49}
\def\ge{G111.23$-$01.23}
\def\gf{G111.25$-$00.76}
\def\j23{J2302$+$6405}
\def\masyr{~mas~yr$^{-1}$}
\def\cepa{Cepheus A}
\def\ngc{NGC\,7538}

\draft
\SetRunningHead{J. O.~Chibueze\etal}{Trigonometric Parallax of \r22 with VERA}

\title{Trigonometric Parallax of \r22 with VERA: 3-Dimensional View of Sources Along the Same Line of Sight}
\author{James O. \textsc{Chibueze}\altaffilmark{1,2},
Hirofumi \textsc{Sakanoue}\altaffilmark{3}, 
Takumi \textsc{Nagayama}\altaffilmark{4},
Toshihiro \textsc{Omodaka}\altaffilmark{3}, \\ 
Toshihiro \textsc{Handa}\altaffilmark{3}, Tatsuya \textsc{Kamezaki}\altaffilmark{3}, 
Ross \textsc{Burns}\altaffilmark{3}, Hideyuki \textsc{Kobayashi}\altaffilmark{4}, Hiroyuki \textsc{Nakanishi}\altaffilmark{3}, Mareki \textsc{Honma}\altaffilmark{4}, Yuji \textsc{Ueno}\altaffilmark{4}, Tomoharu \textsc{Kurayama}\altaffilmark{5}, Mitsuhiro \textsc{Matsuo}\altaffilmark{3}, and Nobuyuki \textsc{Sakai}\altaffilmark{4}}

\altaffiltext{1}{East Asian ALMA Regional Center, National Astronomical Observatory of Japan, 2-21-1 Osawa, Mitaka, Tokyo 181-8588, Japan}
\altaffiltext{2}{Department of Physics and Astronomy, Faculty of Physical Sciences, University of Nigeria, Carver Building, 1 University Road, Nsukka, Nigeria.}
\altaffiltext{3}{Department of Physics and Astronomy, Graduate School of Science and Engineering, Kagoshima University, 1-21-35 Korimoto, Kagoshima 890-0065, Japan.}
\altaffiltext{4}{Mizusawa VLBI Observatory, National Astronomical Observatory of Japan, 2-21-1 Osawa, Mitaka, Tokyo 181-8588, Japan.}
\altaffiltext{5}{Center for Fundamental Education, Teikyo University of Science, 2525 Yatsusawa, Uenohara, Yamanashi 409-0193 Japan.}

\email{james.chibueze@nao.ac.jp}
\KeyWords{Galaxy: kinematics and dynamics --- techniques: interferometric --- stars:individual (\r22)}
\Received{2014/3/?}
\Accepted{2014/5/?}
\Published{2014/10/?}

\begin{document}
\setlength{\baselineskip}{6ex}

\maketitle

\begin{abstract}
We report the results of the measurement of the trigonometric parallax of an H$_2$O maser source in \r22 with the VLBI Exploration of Radio Astrometry (VERA). The annual parallax was determined to be 0.314$\pm$0.070~mas, corresponding to a distance of 3.18$^{+0.90}_{-0.66}$~kpc. Our results confirm \r22 to be located in the Perseus arm. We computed the peculiar motion of \r22 to be $(U_{src}, V_{src}, W_{src}) = (4\pm4, -32\pm6, 8\pm6)$~\kms, where $U_{src}$, $V_{src}$, and $W_{src}$ are directed toward the Galactic center, in the direction of Galactic rotation and toward the Galactic north pole, respectively.
\r22, NGC\,7538 and Cepheus A lie along the same line of sight, and are within 2$^{\circ}$ on the sky. Their parallax distances with which we derived their absolute position in the Milky Way show that \r22 and \ngc~are associated with the Perseus Arm, while \cepa~is located in the Local Arm. We compared the kinematic distances of \iras, \ga, \gb, \gc, \gd, \cepa, \ngc, \ge, \gf~and \r22~derived with flat and non-flat rotation curve with its parallax distance and found the kinematic distance derived from the non-flat rotation assumption ($-$5 to $-$39 \kms~lag) to be consistent with the parallax distance. 
\end{abstract} 

\section{Introduction}
\label{sec:introduction}
Recent large surveys of molecular line emissions towards the Galactic plane (e.g. Dame et al. 2001) have greatly improved our understanding of the spatial distribution, and the kinematics of giant molecular clouds (GMC) in the Milky Way. Based on such surveys, large scale physical properties (like luminosity, column density, temperature, mass) of molecular clouds can be derived, yet, the 3-dimensional distribution of the clouds or components of a GMC remain obscured. This is largely due to the ambiguity in the kinematic distances of the clouds estimated from their local standard of rest velocity (\vlsr), which is dependent on how accurately the \vlsr of the cloud is determined. The kinematic distance is also dependent on the Galactic rotation model based on the Galactic rotation curve and the Galactic constants. Therefore, the independent and accurate measurements of the source distance are important to study 3D structure of Milky Way.

From 107$^{\circ}$ to 112$^{\circ}$~ in the Galactic longitude lies \iras, \ga, \gb, \gc, \gd, \cepa, \ngc, \ge, \gf~(Reid et al. 2014; Hirota et al. 2008; Rygl et al. 2010; Moscadelli et al. 2009; Choi et al. 2014) and \r22, whose parallax distances are known.
Cepheus A, NGC\,7538, and \r22 are located within 2$^{\circ}$ on the sky (Ungerechts et al. 2000), and this may lead to the wrong conclusion that these sources are located at the same distance or in the same GMC. Moscadelli et al. (2009) measured the parallax distance to Cepheus A (see also  Dzib et al. 2011) and NGC\,7538 to 0.70$\pm$0.04 kpc and 2.65$\pm$0.12 kpc, respectively, indicating that Cepheus A is actually located in the Local arm, while NGC\,7538 is located in the Perseus arm.

Compared to NGC\,7538, \r22 looks even spatially closer to Cepheus A from the Galactic plane point of view, but their \vlsr are different. The \vlsr of Cepheus A, NGC\,7538, and \r22 are $-$10, $-$53, and $-$63~\kms, respectively (Ungerechts et al. 2000). Considering the importance of distance in diagnosing the physical conditions/properties such as luminosity, masses, sizes of star forming regions, and studying the evolution of the young stellar and/or proto-stellar objects within them, we have made multi-epoch very long baseline interferometry (VLBI) astrometric observations toward \r22 to determine its accurate distance and compare it with the known parallax distances of Cepheus A and NGC\,7538.

There is no existing information about \r22 in literatures, except that its position (from SIMBAD) being close to Cepheus A star forming region. In our \h2o maser survey observations with the Kagoshima 6~m radio telescope, we detected \h2o masers towards \r22 for the first time. In this paper, we report the annual parallax of the newly detected \h2o masers of \r22.

\section{Observations and data reduction}
\label{sec:observation}

The observations of the H$_2$O (6$_{15}$--5$_{23}$) masers in \r22 at a rest frequency of 22.235080 GHz were carried out in 11 epochs spread over about 3 years from 2010 May 28 to 2013 January 26 using VLBI Exploration of Radio Astrometry (VERA). The observation details of the 11 epochs are summarized in Table \ref{tab:tab1}. Each observation epoch spanned $\sim$ 8 hours, covering scans on \r22, a bandpass calibrators (3C454.3, NRAO530 and DA55), and the position reference source (\j23). The dual-beam system of the VERA telescopes enabled us to simultaneously observe the H$_2$O masers of \r22 and a position reference source (\j23). 
\r22 and \j23 are spatially separated by 1$^{\circ}$.7. 
The data were recorded onto magnetic tapes at the rate of 1024 Mbps, providing a total bandwidth of 256~MHz, which were digitized in four quantization levels, and then divided into 16 base-band channels (BBCs) with a bandwidth of 16~MHz each. One of the BBCs was assigned to the H$_2$O maser emission in \r22, while the remaining 15 BBCs were assigned to the continuum emission from the position reference source.

The data correlation was made with the Mitaka FX correlator. The accumulation period of the correlation was set 
to 1 second. The correlation outputs consisted of 512 and 16 spectral channels for the H$_2$O maser and reference 
continuum emission, respectively. The H$_2$O maser spectral resolution was 31.25 kHz, corresponding to a velocity spacing of 0.42 km~s$^{-1}$.

The astrometric data reduction was carried out with the VERA Data Analyzer (VEDA) developed by the software development group at the Mizusawa VLBI Observatory of the National Astronomical Observatory of Japan for astrometric analysis of VERA data (Niinuma et al. 2011). We adopted the following phase tracking centers for the observed sources;  \j23 (position-reference source) ($\alpha_{J2000}=$ 23$^{\mbox h}$02$^{\mbox m}$41$^{\mbox s}.$315001 and $\delta_{J2000}=~$+$~\!\!$64$^{\circ}$05$^{\prime}$52$\arcsec.$84853),
 ~while the position of \r22 was set as $\alpha_{J2000}=$ 22$^{\mbox h}$57$^{\mbox m}$29$^{\mbox s}.$809 and $\delta_{J2000}=~$+$~\!\!$62$^{\circ}$29$^{\prime}$46$\arcsec.$85,
 ~in a recalculation of the delay tracking model.

The VEDA pipeline data reduction steps include time and frequency averaging (Niinuma et al. 2011) of the position-reference source, \j23, to reduce the size of the dataset. Amplitude and bandpass calibrations were performed with the total-power spectra of the bandpass calibrator. The delay and delay-rate offset were determined by a fringe search process on the used calibrator. The station solutions determined by calibrator were used in the fringe search of \j23, after which the image were obtained by CLEAN deconvolution and then self-calibration was performed on the obtained image. The phase solutions of \j23, obtained with integration time of 2 min, was applied to the maser source to obtain its absolute position. The average of the peak intensity of \j23 in all the epochs was $\sim$ 80 mJy~beam$^{-1}$. 
Due to bad weather conditions, the position reference source was not detected in 3 out of the 11 epochs, and those 3 epochs were not used in the parallax measurement because we are unable to determine the absolute position of the maser spots. Doppler tracking was carried out on the maser source. The velocity channels containing bright maser spots were imaged with a field of view of 2\arcsec $\times$ 2\arcsec~with 1024 $\times$ 1024 grids in order to find maser spots detected at a signal to noise ratio over 5. The final images of individual detected maser spots were obtained each with an area of 10 mas $\times$ 10 mas (size) with 512 $\times$ 512 grids.

\section{Results}
\label{sec:results}

\subsection{Annual Parallax of  \r22}
\label{sec:results1}
\begin{figure*}
\begin{center}
\FigureFile(110mm, 170mm){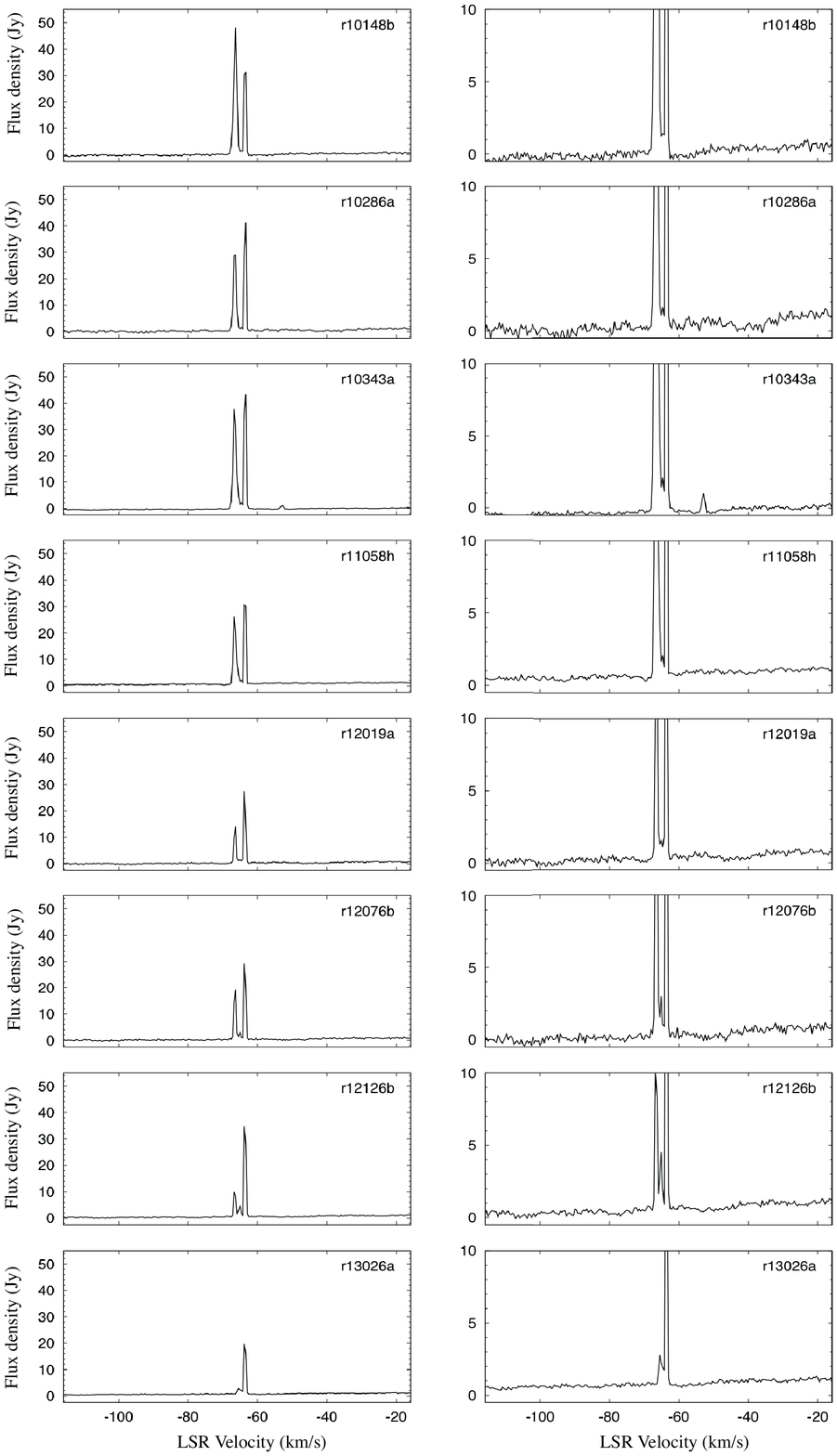}
\end{center}
\caption{Spectra of the \h2o masers at the different epochs of observation.} 
\label{fig:fig2}
\end{figure*}

Figure \ref{fig:fig2} shows the cross power spectra of the \h2o masers in \r22 detected on the Mizusawa-Iriki baseline at the different epochs. The spectra show conspicuous double peaks.
~The \vlsr range of the detected \h2o masers is from $-$ 67.9~km~s$^{-1}$ to $-$ 65.9~km~s$^{-1}$, the strongest maser component detected at $\sim$ 40~Jy. There is a noticeable difference in spectra of the last epoch taken on the 26 January, 2013, which more than half a year after the last usable epoch (5 May, 2012). A total of 39 \h2o maser spots where detected, and we were able measured the proper motion of 15 of the spots. These 15 spots were used to perform a combined fit for the trigonometric parallax of \r22 (see Table \ref{tab:tab2}). Assuming that averaging the internal motions of the maser spots cancel out the influence of external forces on their motions, we derived the systemic proper motion of \r22 to be $-$2.04$\pm$0.05~\masyr, and $-$0.66$\pm$0.06~\masyr~in the R.A. and Declination, respectively (similar procedure can be found in Hachisuka et al. 2008; Nagayama et al. 2011). We added a virial-like term of 5~\kms~ ($\sim$~0.3~\masyr) to the error values of the combined proper motion and \vlsr, the obtained values ($-$2.04$\pm$0.35~\masyr, $-$0.66$\pm$0.36~\masyr, and $-$63$\pm$6~\kms~in the R.A., Declination, and \vlsr, respectively) were used to calculate the Galactic rotation and the peculiar motion.

\begin{figure*}
\begin{center}
\FigureFile(170mm, 110mm){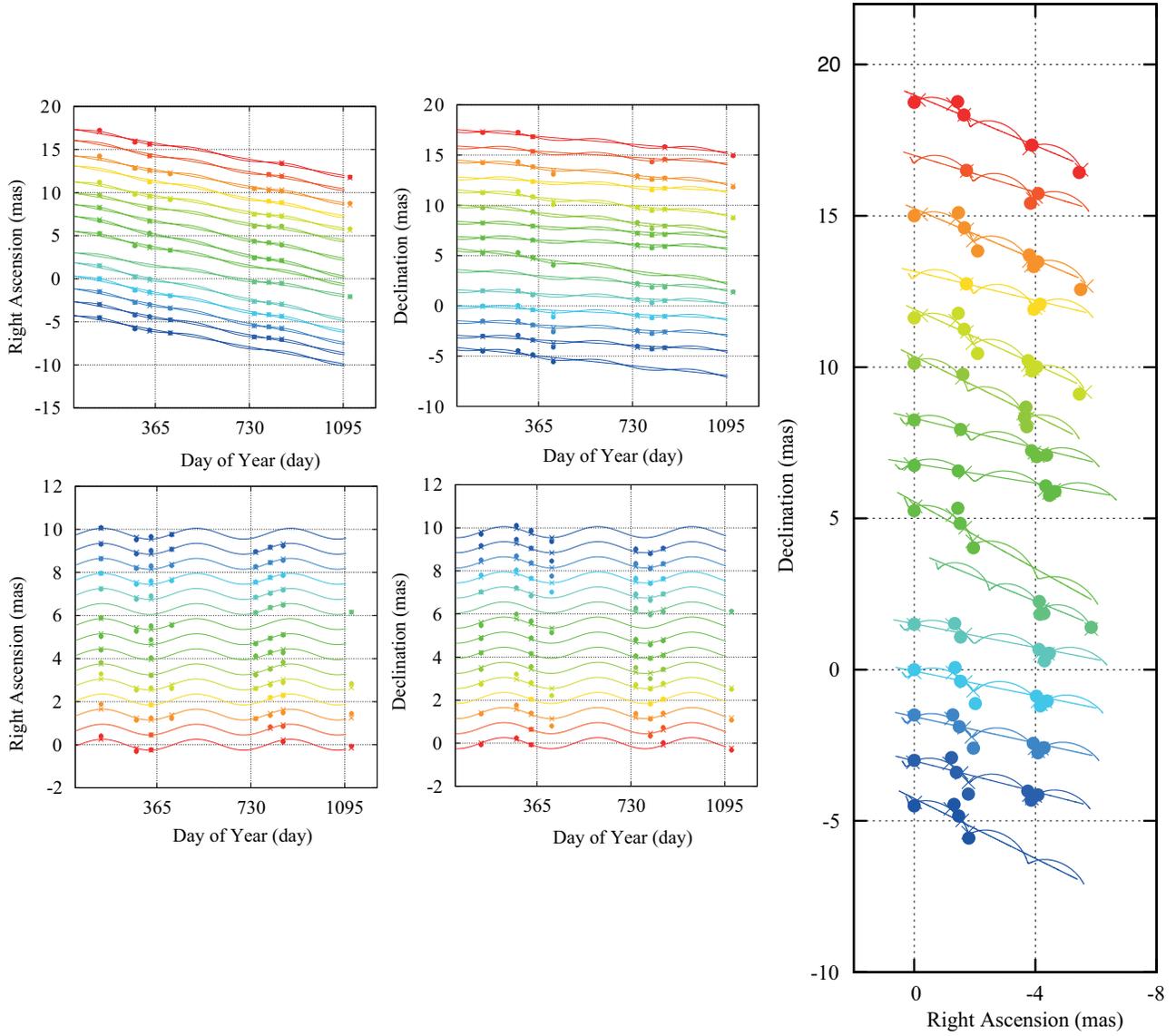}
\end{center}
\caption{Parallax fitting results for the 15 detected \h2o maser spots. {\it Right}: R.A against Declination fits of the parallax and proper motions of the masers. The middle and left panels show the positional change of the \h2o masers at the different epochs, with the proper motion (top) and without the proper motion (bottom). The colors represent the \vlsr of the masers spots (see Figure 4).} 
\label{fig:fig3}
\end{figure*}

Figure \ref{fig:fig3} shows the positions of the maser spot used for the parallax fitting and absolute proper motion fittings. The statistical error in position of the individual maser spots was estimated from the signal-to-noise ratio to range from 0.02 -- 0.08 mas. The positional variations show systematic sinusoidal modulation with a period of 1~year, caused by the parallax.
From the combined fitting, the annual parallax ($\pi$) was obtained to be 0.279$\pm$0.019~mas. Conservatively, we assumed that some of the maser spots are not independent, thus we combined and averaged the parallaxes of spot 3 and 5, as well as spots 11 to 14. We derived the unweighted mean parallaxes (and the corresponding uncertainties) of the 2 combined maser features and the remaining 7 independent maser spots to be 0.314$\pm$0.070~mas, corresponding to a distance of 3.18$^{+0.90}_{-0.66}$~kpc. The standard deviations of the post-fit residuals were 0.127 and 0.183 mas in right ascension and declination, respectively. We introduced an error floor of 0.127 mas and 0.183 mas in right ascension and declination, respectively, for all of the results of the position measurements to make the reduced $\chi^{2}$ to be unity in the least-squares analysis. These error floors can be defined as  the positional uncertainties in the astrometric observations that may have originated from the difference in the optical path lengths between the target and reference sources, caused by the atmospheric zenith delay residual and/or a variability of the structure of the maser spot (Honma et al. 2007; Hirota et al. 2007).


\subsection{Internal Motion of Gas around \r22}
\begin{figure*}
\begin{center}
\FigureFile(170mm, 110mm){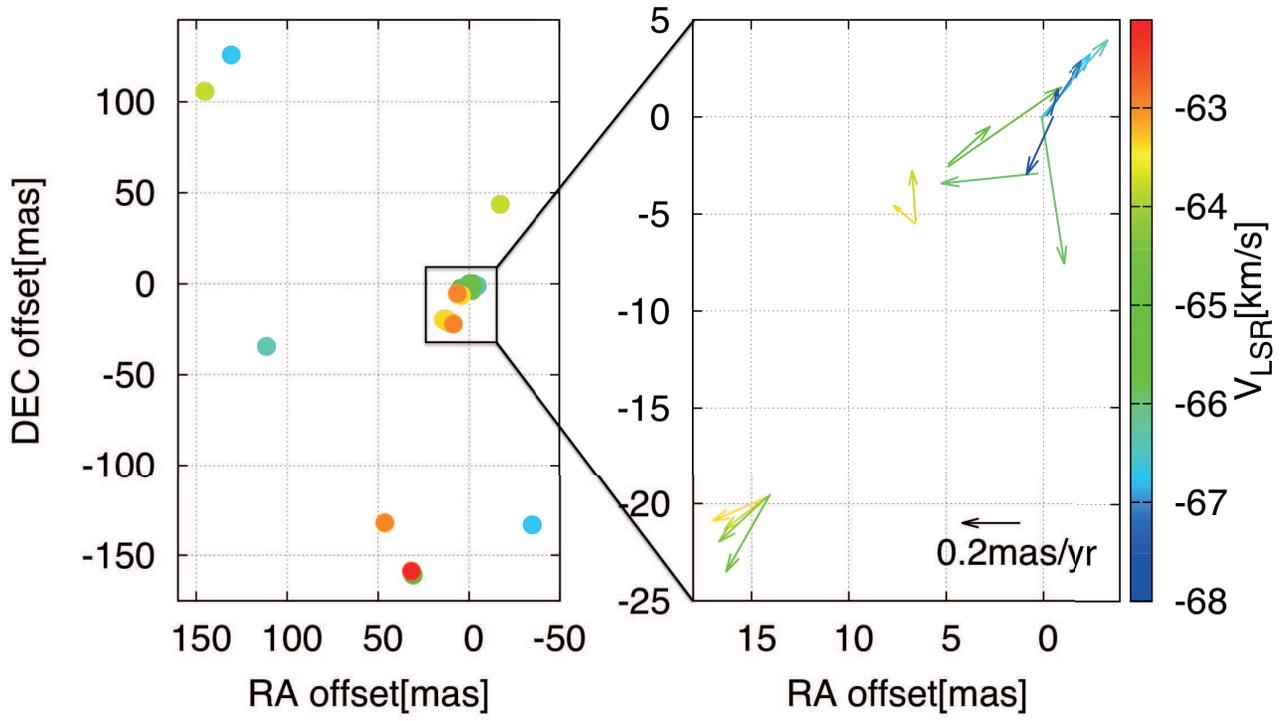}
\end{center}
\caption{{\it left}: The spatial distribution of the \h2o maser spots detected in our observations. {\it Right}: Internal motions of the 15 maser spots. The color codes represent the \vlsr of the masers spots.} 
\label{fig:fig4}
\end{figure*}

Figure \ref{fig:fig4} (left panel) shows the distribution of the 39 \h2o maser spots detected in our observations (see Table \ref{tab:tab2}). The maser spots are distributed over a 200 $\times$ 350 mas (636 $\times$ 1113 AU at the distance of 3.18 kpc) area. This is also a noticeable concentration of about 40\% of the detected maser spots at the center of origin of the maser map (Figure \ref{fig:fig4}). To understand the nature of the gas motion around the driving source, \r22, we traced the absolute proper motion of the maser spots detected in 3 or more epochs. 

Figure \ref{fig:fig4} (right panel) shows the traced internal motion of the masers spots. The average of the obtained internal motions is 0.25~\masyr, corresponding to 3.8 \kms~at 3.18kpc.
The measured proper motions show a northwest-southeast bipolar structure, typical of the bipolar outflow of a massive young stellar object (Chibueze et al. 2012, 2014). This is consistent with the observed double peaks seen in the spectra of the maser emissions (see \ref{sec:results1}). The northwest maser group are dominated by blue-shifted spots, while the southwest group is dominated by the redshifted spots. Assuming a symmetrical bipolar structure, it could be suggested that the driving source, \r22, is located at the center of the main axis of the bipolar outflow. But, the randomness of motion of the northwest maser group could be explained by some turbulence, which may be due to the closeness of the maser group to the YSO.

The peculiar motion of \r22 is dependent on the motion of the Sun, and that of the Milky Way. Assuming that the driving source of the \h2o masers is located within 25~mas from the origin of the maser map (Figure \ref{fig:fig4}), we derived the systemic motion of the source to be the mean deviation of maser proper motions. This cancels out the influence of star formation activities (like outflow, jets, infall) on the maser motion, leaving us with only the motion that is intrinsic to the driving source. The obtained values of the motion is  2.04$\pm$0.35 \masyr~and $-$0.66$\pm$0.36 \masyr~in the Right Ascension and Declination axes, respectively.

We used the \vlsr obtained from the Ungeretchs et al. (2000) CO survey, and the value is $-$63.0$\pm$6.0 km~s$^{-1}$. We converted the positions and velocities of \r22 from the equatorial heliocentric reference frame to the Galactic reference frame following the procedure of Reid et al. (2009).
We assumed a flat rotation of the Galactic frame. The distance of the Sun from the Galactic center ($R_0$) and the rotation speed of the local standard of rest ($\Theta_0$) were adopted as 8.05 kpc and 238~\kms, respectively (Honma et al. 2012). The solar motion was taken from Sch{\"o}nrich, Binney, \& Dehnen (2010) as $(U_{\odot}, V_{\odot}, W_{\odot}) = (11.10\pm0.75, 12.24\pm0.47, 7.25\pm0.37)$ \kms. 

Therefore, we obtained the peculiar motion of \r22 to be $(U_{src}, V_{src}, W_{src}) = (4\pm4, -32\pm6, 8\pm6)$ \kms, where $U_{src}$, $V_{src}$, and $W_{src}$ are directed toward the Galactic center, in the direction of Galactic rotation and toward the Galactic north pole, respectively. 

\section{Discussion}
\label{sec:discussion}
\subsection{Sources with Known Parallaxes within Galactic Longitude 107$^{\circ}$ -- 112$^{\circ}$}
\begin{figure*}
\begin{center}
\FigureFile(170mm, 110mm){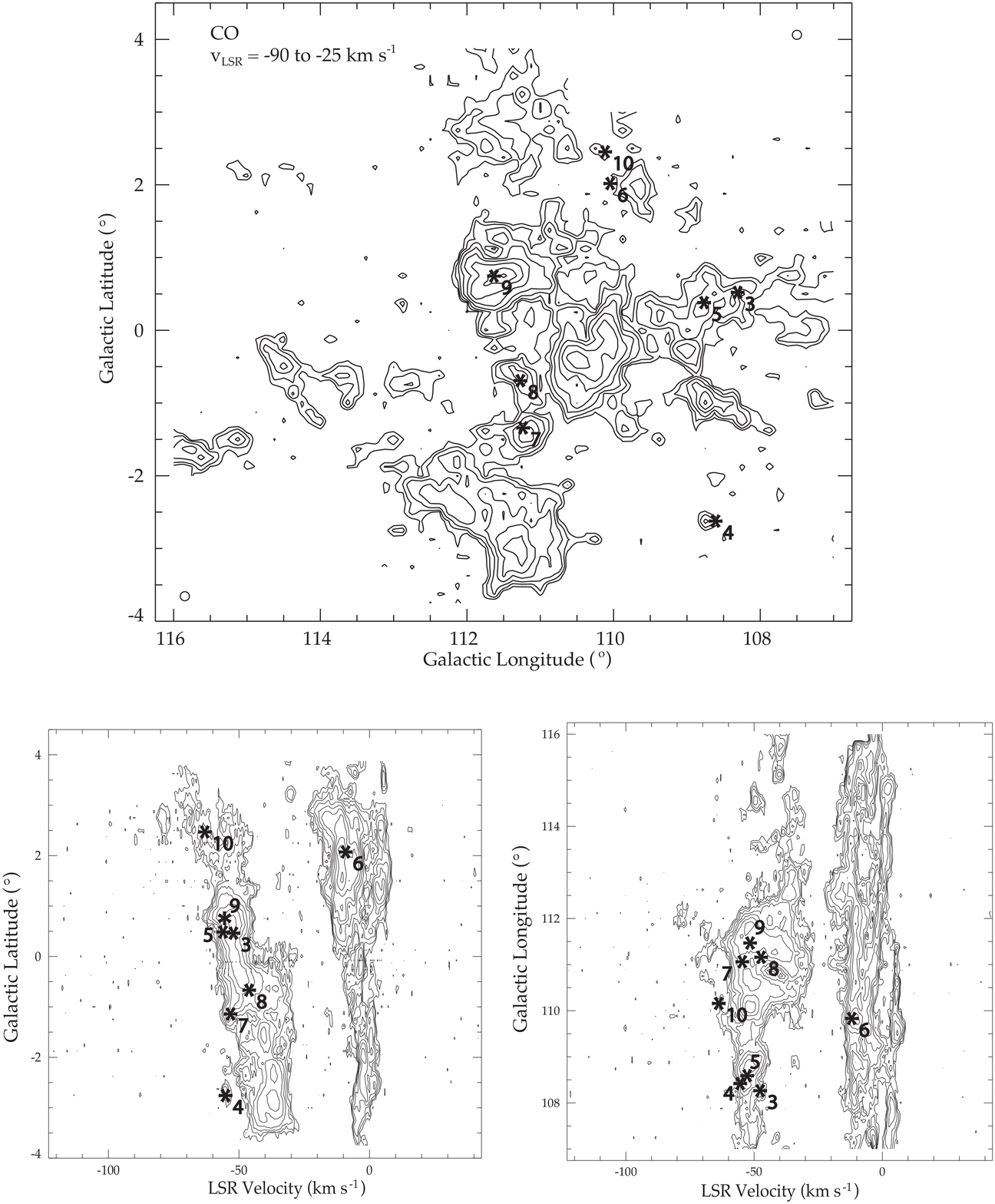}
\end{center}
\caption{Top: $^{12}$CO map of Ungeretchs et al. (2000) showing the positions of source  (represented with $\ast$) 3 to 10, source 1 (\iras) and 2  (\ga) are out of the map range. Sources 3, 4, 5, 6, 7, 8, 9, and 10 represent  \gb, \gc, \gd, \cepa, \ngc, \ge, \gf, and \r22, respectively (see Table 3). The bottom panels show the position-velocity diagrams along the Galactic latitude (bottom left) and the Galactic longitude (bottom right). The symbols and numbers are the same as in the top panel.} 
\label{fig:fig5}
\end{figure*}

Within Galactic longitude~107$^{\circ}$~to~112$^{\circ}$~\iras, \ga, \gb, \gc, \gd, \cepa, \ngc, \ge, \gf~(Reid et al. 2014, and references therein), and \r22 have known parallax measurements.
Figure 4 shows the locations of these sources (except \iras~and \ga) on the $^{12}$CO $l$-$b$ and its position-velocity maps (Ungerechts et al. 2000). These sources are located within 5$^{\circ}$ on the sky (see Figure \ref{fig:fig5} top panel).
The \vlsr~of the sources varies from $-$10 \kms to $-$63 \kms (see Table 3). There is a difference in \vlsr of $\sim$50 \kms~between \r22 and Cepheus A, $\sim$10 \kms~between NGC\,7538 and \r22, $\sim$40 \kms~between NGC\,7538 and Cepheus A. The measurement of the parallax distance of \r22 has enabled us to clearly identify the location of the source and to compare it with location of the rest of the sources. The parallax distance of \r22 differs with those of Cepheus A and NGC\,7538 (Moscadelli et al. 2009) by $\sim$3 kpc and $\sim$1 kpc respectively. While \iras, \ga, and \cepa~ are located in the Local arm, \gb, \gc, \gd, \cepa, \ngc, \gf, and \r22~are located on the Perseus arm.

\begin{figure*}
\begin{center}
\FigureFile(170mm, 110mm){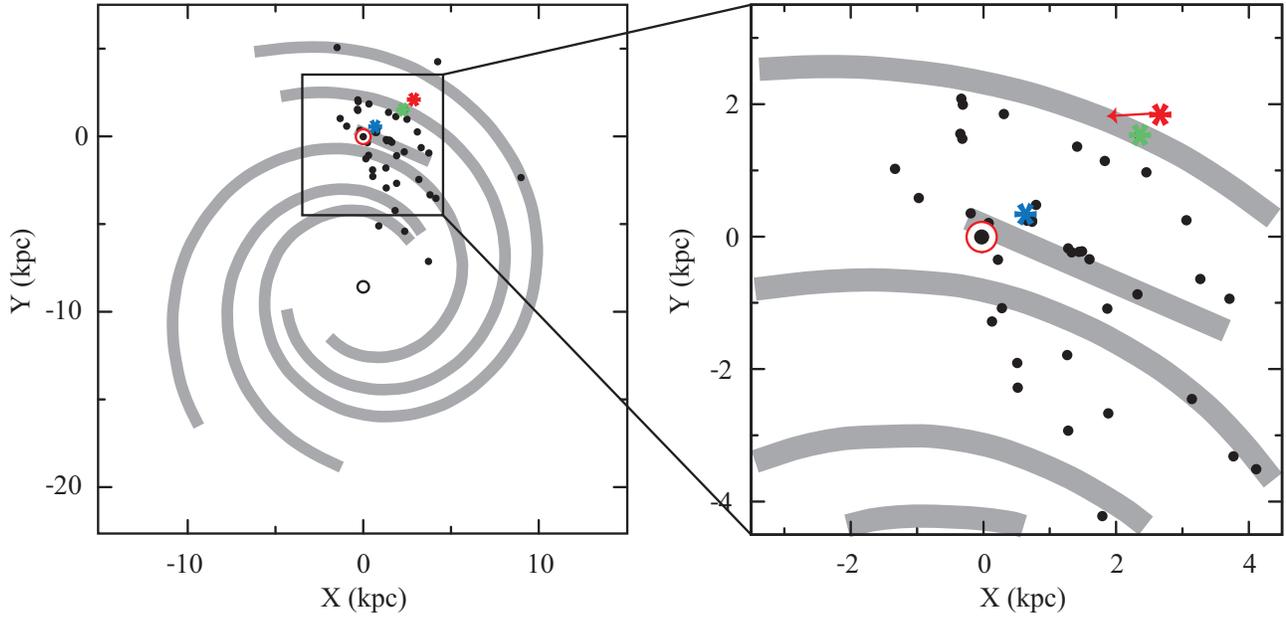}
\end{center}
\caption{Position of the Cepheus A (blue star), NGC\,7538 (green star), and \r22 (red star) on the Galactic plane with the spiral arms based on the Russeil (2003) arm model. Positions of the sources (black filled circles) reported by Honma et al (2012) on Galactic plane, the Sun is represented by a black filled circle enclosed in a red circle. The arrow indicates the peculiar motion of \r22.} 
\label{fig:fig8}
\end{figure*}

Figure \ref{fig:fig8} shows the location of \cepa, \ngc, and \r22~(from their parallax distances) in the Milky Way based on the arm model of Russeil (2003). We have plotted only \cepa, \ngc, and \r22~for clarity.

\subsection{Rotation Curve and Kinematic Distance}

Flat rotation is usually assumed in the estimation of the kinematic distance to sources in the Milky Way (Reid et al. 2009). Such assumption implies that the rotation curve of the Milky Way is flat, thus, the ratio of the change in the rotation speed of the Milky Way to the change of the distance of a source from the Galactic center will be zero. Sofue et al. (2009) presented a non-flat unified rotation curve of the Milky Way using  existing data (Sofue et al. 2009; and references therein).
Therefore, the accuracy of the estimation of the kinematic distance of a source from its \vlsr is dependent on the Galactic constants, assumed rotation curve of the Milky Way and the accuracy of the \vlsr.

\begin{figure*}
\begin{center}
\FigureFile(170mm, 110mm){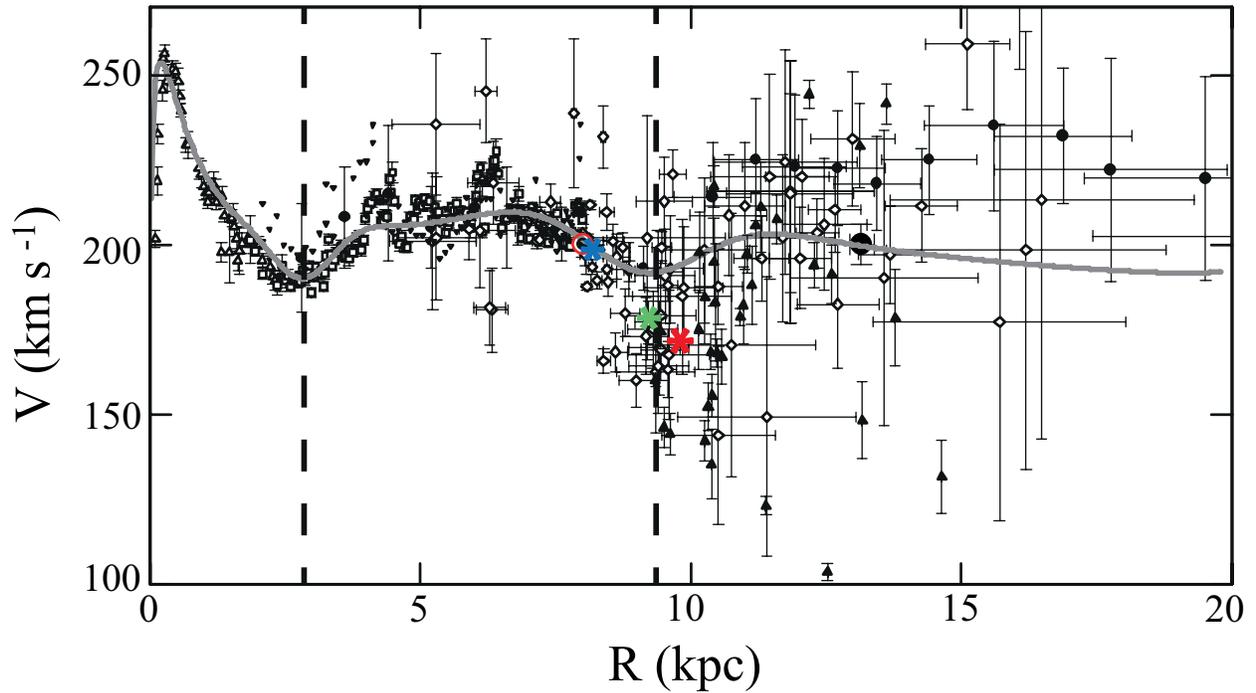}
\end{center}
\caption{Rotation curve (adopted from Sofue et al. 2009) showing the plot of the distance from the Galactic center (R) against the rotation velocity (V). The positions of the Sun (black filled circle enclosed in a red circle), Cepheus A (blue star), NGC\,7538 (green star), and \r22 (red star) are shown in the plot. The dash lines indicate the dips at 3 and 9 kpc.} 
\label{fig:fig9}
\end{figure*}

Using the Galactic constants of Honma et al. (2012) (Galactic rotation, $\Theta_0$ of 238 \kms), and adopting solar motion of Sch{\"o}nrich et al. (2010) $(U_{\odot}, V_{\odot}, W_{\odot}) = (11.1, 12.24, 7.25)$ \kms, we calculated the kinematic distances of \cepa, \ngc, and \r22 using Reid et al. (2009) code. We obtained the kinematic distances both for the case of flat rotation assumption and the also using the rotation curve of Sofue et al. (2009). Table \ref{tab:tab3} shows the kinematic distances of the sources in both cases and their parallax distances. 
Figure \ref{fig:fig9} shows the rotation curve of Milky Way Galaxy including IRAS22555+6213, Cepheus A, and NGC7538 (only 3 sources for clarity). \r22 has a lag of $\sim -$39~\kms in the Galactic rotation compare with that determined with the Sun (adopted from Sofue et al. 2009). 
With respect to the rotation velocity of LSR at the Galactocentric distances of the sources, we found a lag ranging from $-$5 to $-$39 \kms~(see Table 3), using their parallax distances as benchmarks. 

Sofue et al. (2009) reported two prominent perturbation dips at 3 and 9 kpc from the Galactic center. Figure 6 shows 2 deep minima at 3 and 9 kpc. The 3-kpc dip was attributed to the bar observed with the Cosmic Background Explorer (COBE), which could be related to the ring of radius 4 kpc. The 9-kpc dip was also confirmed in their results. They suggested that a massive ring at 11 kpc is responsible for the observed dip at 9 kpc. Our parallax measurement of \r22 and the consistency of the kinematic distance of the source obtained from the non-flat rotation curve (Sofue et al. 2009), coincides with the 9-kpc dip. This explains the over $-$20 \kms~lags seen in the rotation velocity of  \gc, \gd, \ge, \ngc, and \r22. Surprisingly, two of the Perseus arm sources, \gb, and \gf, have lags comparable to those in the Local arm. Our results support the existence of the 9-kpc dip (Sofue et al. 2009; Bhattacharjee et al. 2014). We cannot confirm if the 9-kpc dip is a global effect around the entire Milky Way, but it is dominated by sources in the Perseus arm between Galactic longitude 110$^{\circ}$ to 135$^{\circ}$. The parallax and proper motion measurements of many sources in this region indicate significant peculiar motions largely counter to Galactic rotation (Xu et al 2006). This and the presence of shorter lag sources may suggest that the 9-kpc dip may rather be a localized  phenomenon in the Perseus arm, and not a global effect around the entire Galaxy.


\section{Conclusions and Summary}

We report the results of the trigonometric parallax measurement of newly detected \h2o masers associated with \r22, whose annual parallax was determined to be 0.314$\pm$0.070~mas, corresponding to a distance of 3.18$^{+0.90}_{-0.66}$~kpc. We obtained the peculiar motion of \r22 to be $(U_{src}, V_{src}, W_{src}) = (4\pm4, -32\pm6, 8\pm6)$ \kms.

On the Galactic Plane, \cepa, \ngc ~and \r22 lie along line-of-sight, but are located at different positions in the Milky Way. While \cepa~is associated with the Local Arm, \ngc~and \r22 are located in the Perseus Arm.

The accuracy of derived kinematic distance of source is dependent of the assumed rotation of the Milky Way. The kinematic distance of \r22~obtained based on the assumption of flat rotation has a large error percentage, while that one obtained with non-flat unified rotation curve of Sofue et al. (2009) is consistent with the parallax distance of the source.

\bigskip
We acknowledge all VERA staff members and students who have helped with the array operation and with the data correlation.

\clearpage

\clearpage

\clearpage

\begin{table*}[t]

\begin{center}

\begin{minipage}{0.9\textwidth}

\caption{VERA observations of \r22.}
\label{tab:tab1}
\footnotesize

\begin{tabular}{cccccccc} 
\hline

& Observation{\phantom{aaa}}  &  &  J2302$-$6405 & & VERA{\footnotemark[a]}$T_{\rm sys}$  &  \\

{Code} & {Epoch} &  Astrometry? & Peak Intensity & MIZ & IR & OG & ISH\\ 
 &  &  & (Jy~beam$^{-1}$) & (K) & (K) & (K) & (K) \\

\hline 
r10148b & 2010 May 28 & Yes & 0.103 & 162.9 & 214.4 & 298.3 & 844.2\\
r10286a & 2010 October 13 & Yes & 0.096 & 327.6 & 264.1 & 383.7 & 606.7\\
r10343a & 2010 December 9 & Yes & 0.058 & 173.6 & 115.0 & 172.6 & 174.9 \\ 
r11058h & 2011 February 27 & Yes & 0.076 & 177.7 & 391.8 & 167.9 & 233.5 \\ 
r11214a & 2011 August 2 & No & --  & 427.9 & 365.9 & 596.3 & 390.2\\ 
r11309a & 2011 November 5 & No & -- & 474.5 & 1100.4 & 235.5 & 385.4 \\ 
r12019a & 2012 January 19 & Yes & 0.085 & 118.2 & 1091.9 & 192.1 & 240.6 \\ 
r12076b & 2011 March 16 & Yes & 0.094 & 253.1 & 518.3 & 192.1 & 323.7\\ 
r12126b & 2012 May 5 & Yes & 0.071 & 186.3 & 152.5 & 233.5 & 268.5\\ 
r12232a & 2012 August 19 & No & -- & 275.8 & 445.8 & 869.9 & 565.8\\ 
r13026a & 2013 January 26 & Yes & 0.063 & 183.7 & 173.4 & 150.5 & 201.2\\
\hline 
\end{tabular}

\footnotetext[a]{VERA comprises of Mizusawa (MIZ), Iriki (IR), Ogasawara (OG), and Ishigakijima (ISH) stations, and all the stations participated in all the observation epochs.}
\end{minipage}
\end{center}
\end{table*}

\begin{table*}[t]

\begin{center}

\begin{minipage}{1.0\textwidth}

\caption{Annual Parallax Fitting Results and the Internal Motions of the \h2o Masers Associated with \r22).}
\label{tab:tab2}
\footnotesize

\begin{tabular}{cccccccc} 
\hline 

{Feature} & {\vlsr} &  $\Delta\alpha$ cos$\delta$ & $\Delta\delta$ & Epochs & $\pi$ & $\mu_{\alpha}$cos$\delta$ & $\mu_{\delta}$ \\
ID & (\kms) & (mas) & (mas) &  & (mas) & (\masyr) & (\masyr) \\

\hline 
1 &$-$63.372 &14.214 &$-$19.666 &123***78 & 0.358$\pm$0.079 &$-$1.86$\pm$0.07 & $-$0.74$\pm$0.05\\
2 &$-$63.372 &6.558 &$-$5.524 &**3**67* & -- &$-$1.97$\pm$0.09 & $-$0.60$\pm$0.16 \\
3 &$-$63.793 &14.220 &$-$19.681& 12345678 & 0.351$\pm$0.085 &$-$1.90$\pm$0.08 &$-$0.77$\pm$0.08\\
4 &$-$63.793 &6.554 &$-$5.395 & **3**67* & -- &$-$2.03$\pm$0.04 &$-$0.49$\pm$0.24 \\
5 &$-$64.215 &14.175 &$-$19.631 &12345678 & 0.378$\pm$0.084 &$-$1.88$\pm$0.08 &$-$0.82$\pm$0.07 \\
6 &$-$64.636 &14.039 &$-$19.518 &1*3*567* & 0.356$\pm$0.039 &$-$1.89$\pm$0.07 &$-$0.93$\pm$0.12 \\
7 &$-$65.057 &4.895 &$-$2.432 &1*3*567* & 0.161$\pm$0.036 &$-$2.19$\pm$0.07 &$-$0.53$\pm$0.12 \\
8 &$-$65.479 &4.914 &$-$2.573 &1*3*567* & 0.155$\pm$0.064 &$-$2.43$\pm$0.07 &$-$0.39$\pm$0.12 \\
9 &$-$65.900 &0.099 &0.037 &1234**** & 0.592$\pm$0.135 &$-$2.12$\pm$0.21 &$-$1.17$\pm$0.36 \\
10 &$-$65.900 &0.260 &$-$2.902 &****5678 & 0.243$\pm$0.021 &$-$1.71$\pm$0.03 &$-$0.70$\pm$0.26 \\
11 &$-$66.321 &0.126 &$-$0.017 &1234567* & 0.225$\pm$0.044 &$-$2.28$\pm$0.06 &$-$0.40$\pm$0.11 \\
12{\footnotemark[a]} &$-$66.743& 0.000& 0.000 &1234567* & 0.243$\pm$0.051 &$-$2.21$\pm$0.06 &$-$0.45$\pm$0.11\\
13 &$-$67.164 &$-$0.147 &0.039 &1234567* & 0.221$\pm$0.042 &$-$2.17$\pm$0.06 &$-$0.47$\pm$0.11 \\
14 &$-$67.585 &$-$0.373 &0.071 &1234567* & 0.211$\pm$0.046 &$-$2.07$\pm$0.06 &0.52$\pm$0.11 \\
15 &$-$68.007 &$-$0.510 &0.043 &1234**** & 0.375$\pm$0.125 &$-$1.95$\pm$0.21 &$-$0.96$\pm$0.36 \\
\hline 
Combine & & & & & 0.279$\pm$0.019 & & \\
Average & & & & & &$-$2.04$\pm$0.05 &$-$0.66$\pm$0.06\\

\hline 
\end{tabular}

\footnotetext[a]{The absolute position of the maser feature ID 12 at the map origin is $\alpha_{J2000}=$ 22$^{\mbox h}$57$^{\mbox m}$29$^{\mbox s}.$806 and $\delta_{J2000}=~$+$~\!\!$62$^{\circ}$29$^{\prime}$46$\arcsec.$837}
\end{minipage}
\end{center}
\end{table*}


\begin{landscape}

\begin{table*}[t]
\begin{center}

\begin{minipage}{0.9\textwidth}

\caption{Comparison of Distances and the Lags of Sources within Galactic Longitude 107$^{\circ}$~to~112$^{\circ}$.}
\label{tab:tab3}
\footnotesize
\scriptsize

\begin{tabular}{cccccccccc} 
\hline 

&& Galactic&Galactic& \vlsr & Kinematic & Kinematic & Parallax  & Lag & Spiral\\
ID{\footnotemark[i]} & Source& Longitude ($^{\circ}$) & Latitude ($^{\circ}$)& (km~s$^{-1}$) & distance{\footnotemark[ii]} (kpc) &distance{\footnotemark[iii]} (kpc) & distance (kpc) & \kms & Arm \\
\hline 
1&\iras & 107.30 & 5.64 & $-11\pm5$& 1.45$\pm$0.67 & 0.75$\pm$0.75 & 0.76$\pm$0.03 &$-$8 & Local \\
2&\ga & 108.18 & 5.52 & $-11\pm3$& 1.42$\pm$0.65 & 0.73$\pm$0.73 & 0.78$\pm$0.10&$-$8 & Local\\
3&\gb & 108.21 & 0.59 & $-49\pm5$ & 4.69$\pm$0.64 & 4.36$\pm$0.64 & 4.37$\pm$..{\footnotemark[iv]} &$-$5 & Perseus\\
4&\gc & 108.47&  $-$2.82 & $-54\pm5$ & 5.12$\pm$0.66 &3.24$\pm$0.68 & 3.24$\pm$..{\footnotemark[iv]} & $-$27 & Perseus\\
5&\gd & 108.60 & 0.49 & $-52\pm5$ & 4.94$\pm$0.65 & 2.49$\pm$0.70& 2.51$\pm$..{\footnotemark[iv]} & $-$34 & Perseus\\
6&\cepa & 109.87 & 2.11& $-10\pm$1 & 1.26$\pm$0.64& 0.68$\pm$0.65 & 0.70$\pm$0.04 & $-$7 & Local \\
7&\ge & 111.24 & $-$1.24 & $-53\pm10$& 4.89$\pm$0.66 & 3.41$\pm$0.66 & 3.47$\pm$..{\footnotemark[iv]} & $-$22 &Perseus\\
8&\gf & 111.26 & $-$0.77 & $-43\pm5$ & 3.99$\pm$0.62 & 3.40$\pm$0.63 & 3.40$\pm$..{\footnotemark[iv]} & $-$9 & Perseus\\
9&\ngc& 111.54& 0.78 & $-53\pm$5 & 4.87$\pm$0.66& 2.66$\pm$0.68 &  2.65$\pm$0.12 & $-$32 & Perseus\\
10&\r22 & 110.20 & 2.48 & $-63\pm$6 &5.89$\pm$0.71 & 3.13$\pm$0.71 &  3.18$\pm$0.90  &  $-$39 & Perseus\\

\hline 
\end{tabular}

\footnotetext[i]{References of the parallax measurements are 1 (Hirota et al. 2008); 2 (Rygl et al. 2010); 3, 4, 5, 7, and 8 (Choi et al. 2014); 6 and 9 (Moscadelli et al. 2009); 10 (this paper); (see also Honman et al. 2012; Reid et al. 2014).}
\footnotetext[ii]{Based on the assumption of flat rotation.}
\footnotetext[iii]{Based on the rotation curve of Sofue et al. (2009).}
\footnotetext[iv]{The exact error values to be published in Choi et al. (2014).}
\end{minipage}
\end{center}
\end{table*}
\end{landscape}

\end{document}